\documentclass[11pt,a4paper]{article}
\pdfoutput=1

\usepackage{bm}
\usepackage{amsmath,amssymb}
\usepackage{cite}
\usepackage{bbold}

\usepackage{graphicx}
\usepackage{color}

\usepackage{hyperref} 
\hypersetup{colorlinks=true, citecolor=blue, filecolor=black, linkcolor=blue, urlcolor=blue, pdfpagemode=UseNone}

\numberwithin{figure}{section}
\numberwithin{equation}{section}

\newcommand{\be}{\begin{equation}}
\newcommand{\ee}{\end{equation}}
\newcommand{\bea}{\begin{eqnarray}}
\newcommand{\eea}{\end{eqnarray}}
\def\beal#1\eeal{\begin{align}#1\end{align}}   
\def\besp#1\eesp{\begin{multline}#1\end{multline}} 

\usepackage[normalem]{ulem}

\newcommand{\kevin}[1]{}

\newcommand{\cL}{\mathcal{L}}

\newcommand\ie{\textit{i.e.}\ }
\newcommand\eg{\textit{e.g.}\ }
\newcommand\cf{\textit{cf.}\ }

\newcommand{\aka}{{a.k.a.}\ }

\newcommand{\half}{\tfrac{1}{2}}

\newcommand{\prop}{\triangle}
\newcommand{\g}{\mathrm{g}}
\newcommand{\one}{\mathbb{1}}

\def\beq{\begin{equation}}
\def\eeq{\end{equation}}

\def\Tr{{\rm Tr}}
\def\tr{{\rm tr}}

\def\str{{\rm str}}

\def\A{\mathcal{A}}

\def\D{\mathcal{D}}
\def\C{\mathcal{C}}

\def\eq#1{(\ref{#1})}

\def\s0#1#2{\mbox{\small{$ \frac{#1}{#2} $}}}
\def\0#1#2{\frac{#1}{#2}}

\def\grgl{\:\hbox to -0.2pt{\lower2.5pt\hbox{$\sim$}\hss}{\raise3pt\hbox{$>$}}\:}
\def\klgl{\:\hbox to -0.2pt{\lower2.5pt\hbox{$\sim$}\hss}{\raise3pt\hbox{$<$}}\:}

\def\hS{\hat{S}}
\def\ldl{\Lambda \partial_{\Lambda}}
\def\dDelta{\dot{\triangle}}
\def\A{{\cal A}}
\def\C{{\cal C}}
\def\gap{\hspace{0.05in}}
\def\ct{\tilde{c}}

\begin{document}

\begin{titlepage}

\begin{center}
{\huge \bf Conformal anomaly 
from gauge fields without gauge fixing} 
\end{center}
\vskip1cm


\begin{center}
{\bf Kevin Falls$^a$ and Tim R. Morris$^b$}
\end{center}

\begin{center}
{$^a$ \it Instit\"{u}t f\"{u}r Theoretische Physik, Universit\"{a}t Heidelberg,\\ Philosophenweg 16, 69120 Heidelberg, Germany}\\
{$^b$ \it STAG Research Centre \& Department of Physics and Astronomy,\\  University of Southampton,
Highfield, Southampton, SO17 1BJ, U.K.}\\
\vspace*{0.3cm}
{\tt Kevin.G.Falls@gmail.com, T.R.Morris@soton.ac.uk}
\end{center}

\abstract{We show how the Weyl anomaly generated by gauge fields, can be computed from manifestly gauge invariant and diffeomorphism invariant exact renormalization group equations, without having to fix the gauge at any stage. Regularisation is provided by covariant higher derivatives and by embedding the Maxwell field into a spontaneously broken $U(1|1)$ supergauge theory. We first provide a realisation that leaves behind two versions of the original $U(1)$ gauge field, and then construct a manifestly  $U(1|1)$ supergauge invariant flow equation which leaves behind only the original Maxwell field in the spontaneously broken regime.}

\end{titlepage}

\tableofcontents

\newpage

\section{Introduction}
\label{sec:Intro}

\kevin{Kevin: please add any further appropriate references either to your own or other people's work - thanks!}

Over a period of some years, a framework has been developed for gauge theory which allows continuum computations without fixing the gauge. This is achieved by utilising the freedom to design manifestly gauge invariant versions of the continuum realisation of Wilson's renormalization group (christened exact RG in ref. \cite{Wilson:1973}). Such manifest gauge invariance was first incorporated into the exact RG in ref. \cite{Morris:1995he}, however in the limited context of pure $U(1)$ gauge theory.  Following ref. \cite{Morris:1998kz} it was generalised and extensively studied first for $SU(N)$ Yang-Mills theory, then QCD \cite{Morris:2006in} and QED \cite{Arnone:2005vd,Rosten:2008zp}. For these gauge theories, regularisation is based on gauge-invariant higher derivatives set at some ultraviolet cutoff scale $\Lambda$, supplemented by gauge invariant Pauli-Villars fields \cite{Morris:1999px} with particular flavours and interactions so that their regularisation properties are preserved under RG flow. It was later realised that the resulting structure could be simply understood as arising from spontaneously broken $SU(N|N)$ super-Yang-Mills theory \cite{Morris:2000fs,Morris:2000jj}. In this scheme, the original gauge field $A^1_\mu$ is joined by a copy gauge field $A^2_\mu$ with wrong sign action, and a complex fermionic (\ie wrong-statistics) gauge field $B_\mu$:
\beq
\label{A}
\mathcal{A}_\mu =   \begin{pmatrix}
A^1_\mu \, \, & \, \, B_\mu\, \\
\bar{B}_\mu \,  & A^2_\mu \\
 \end{pmatrix}  \eeq
This extra regularisation works because these degrees of freedom cancel each other, as happens with Parisi-Sourlas supersymmetry \cite{Parisi:1979ka}, at least sufficiently that, together with appropriately chosen covariant cutoff functions, the theory is then regularised to all orders in perturbation theory \cite{Arnone:2000bv, Arnone:2000qd,Arnone:2001iy}.
The symmetry is then broken spontaneously along the fermionic directions, endowing the $B_\mu$ with a mass at the cutoff scale $\Lambda$. 
The computational methods were generalised in refs. \cite{Arnone:2002yh,Arnone:2002qi,Arnone:2002fa,Arnone:2003pa,Arnone:2002cs} so that universal results could be extracted in a way which was manifestly independent of the detailed form of the regularisation structure, and such that general group invariants could be handled \cite{Arnone:2005fb}. Using these techniques, the initial computation of the one-loop $\beta$ function at infinite $N$ \cite{Morris:1998kz} was generalised to finite $N$ \cite{Arnone:2002qi,Arnone:2002fb,Gatti:2002kc,Arnone:2002cs}, then to two loops \cite{Morris:2005tv,Rosten:2004aw,Arnone:2005fb,Rosten:2005qs,Rosten:2005ka}, extended to all loops in refs. \cite{Rosten:2005ep,Rosten:2006tk} and to computation of gauge invariant operators in refs. \cite{Rosten:2006qx,Rosten:2006pd}. For reviews and further advances see refs. \cite{Arnone:2006ie,Rosten:2010vm,Rosten:2011ty}.

In ref. \cite{Morris:2016nda} the first steps were made in generalising these ideas so as to yield a manifestly diffeomorphism invariant exact RG for use in quantum gravity.\footnote{For an alternative attempt, see ref. \cite{Wetterich:2016ewc}.}
On the one hand the renormalization group structure of quantum gravity is surely of importance \cite{Stelle:1976gc,Adler:1982ri,Weinberg:1980,Reuter:1996}
and on the other hand one can expect conceptual and computational advances from a  framework 
which allows computations to be done while
keeping exact diffeomorphism invariance at every stage, \ie without gauge fixing. Indeed as shown in ref. \cite{Morris:2016nda}, it turns out that these computations can then be done without first choosing the space-time manifold and in particular without introducing a separate background metric dependence. A solution to the difficult issue of background independence is thus automatic in this formalism.\footnote{For a discussion of this issue in the asymptotic safety literature see \eg refs. \cite{Reuter:2009kq,Becker:2014qya,Dietz:2015owa,Labus:2016lkh,Morris:2016spn,Percacci:2016arh,Ohta:2017dsq}.}  However, only the flow equation for classical gravity was developed in ref. \cite{Morris:2016nda}. In this paper we make the first step towards including manifestly diffeomorphism invariant quantum effects involving gravity. 


We will be concerned with the conformal, \aka Weyl or trace, anomaly  \cite{Capper1974b,Duff1977,Duff1994}\footnote{The Weyl anomaly has been investigated in flow equations in refs. \cite{Machado:2009ph,Codello2013,Codello:2015ana}.} generated by gauge fields.  Although this does not involve dynamical gravity, it is clearly important to understand how the known universal answer can arise in this framework, \ie such that gauge invariance is maintained at all stages. Indeed, since the conformal anomaly can be read off from the logarithmically divergent curvature-squared terms at one loop, it is proportional to the signed number of fields (\ie with fermionic fields appearing with opposite sign). Thus in the usual calculation  \cite{Capper1974b,Duff1977,Duff1994}, the ghost degrees of freedom are indispensable. The manifestly gauge invariant formalism reviewed above, proceeds without ghosts. Thus the question arises: working on a curved spacetime, is gauge fixing now necessary to recover the correct Weyl anomaly or not? As we will see in fact the correct Weyl anomaly is reproduced without gauge fixing. This is thus a dramatic confirmation of a formalism that was developed and tested only in flat space calculations. Needless to say, it is also the first time that a manifestly gauge invariant computation has been achieved for the gauge field contribution to the conformal anomaly.  \kevin{As a bi-product we will also compute, without gauge fixing, the one loop gauge field contribution to the (non-universal) divergent corrections to Newton's constant and the cosmological constant.}

For this exercise it is sufficient to consider Maxwell theory, \ie free $U(1)$ gauge fields. As already mentioned, manifest gauge invariance can be straightforwardly incorporated in flow equations for pure $U(1)$ gauge theory in flat space \cite{Morris:1995he}, where in fact only the gauge field $A^1_\mu$ appears. Even for manifestly gauge invariant QED, the gauge field degrees of freedom are not altered or supplemented: only the Dirac fields need regularisation with opposite statistics Pauli-Villars partners \cite{Arnone:2005vd,Rosten:2008zp}. This is because it is straightforward to regularise a $U(1)$ gauge field gauge invariantly using only a cutoff profile $c$ which is a function of partial derivatives rather than covariant derivatives:
\be 
\cL = \frac14 F^1_{\mu\nu}\, c^{-1}(-\partial^2/\Lambda^2) F^{1\,\mu\nu}\,.
\ee
(Throughout we will be working with Euclidean signature.) However the arguments above already show that such a framework could not possibly give the correct Weyl anomaly. In fact once we use a non-flat metric (and thus replace the partial derivatives in $c^{-1}$ with covariant derivatives) we introduce interactions with the metric which destroy the regularisation, since this is then again effectively covariant higher derivative regularisation, which is known to fail at one loop \cite{Slavnov:1972sq,Lee:1972fj,Morris:2016nda}. Therefore even for pure Maxwell theory, we need a wrong-statistics counterpart to play the r\^ole of the Pauli-Villars field. Following the same chain of reasoning as reviewed above, in order for this to be embedded in an exact RG framework, we are led to developing versions of spontaneously broken $U(1|1)$ theory for this purpose. 

As we will see the wrong-statistics fields that are introduced then ensure the correct Weyl anomaly. In fact the result can be directly compared to a more conventional calculation,  although only after rearranging contributions from the wrong-statistics vector and Goldstone fields, reflecting the fact that the supergauge invariance, while spontaneously broken, is nevertheless manifest throughout.

Actually, a wholesale adaptation of the previously developed manifestly gauge invariant methods is not quite what we want, because the $A^2_\mu$ sector is left unbroken. For the purposes of flat space computations
in non-Abelian Yang-Mills theory, this is not a problem \cite{Appelquist:1974tg,Symanzik:1973vg,Arnone:2000bv, Arnone:2000qd,Arnone:2001iy} because all interactions with this sector are irrelevant, starting with
\be 
\tr F^1_{\alpha\beta} F^1_{\gamma\delta}\, \tr F^2_{\epsilon\zeta} F^2_{\eta\theta}
\ee
(with indices contracted in some way), and thus the $A^2$ sector decouples in the continuum limit, providing we work in $D\le4$ dimensions \cite{Arnone:2001iy}. For a computation of the conformal anomaly however, and more generally the purely gravitational action at one loop, the $A^2$ sector will also contribute and thus we expect to find twice the right answer whatever the space-time dimension.\footnote{The wrong sign in the action does not contribute to the metric dependence at one loop.} We will confirm that this is indeed the case. 


While the above framework is enough to work out the Yang-Mills contribution to the pure gravitational action at one loop, its use would clearly be limited beyond this while the unphysical $A^2$ sector remains as part of the continuum theory. We therefore also build an alternative formulation with spontaneous breaking of both the $B_\mu$ and the $A^2_\mu$ field so that all these fields gain masses at the regularisation scale $\Lambda$. This thus leaves only the original unbroken Maxwell field $A^1_\mu$ at low energies,
which, as we will see, then gives exactly the correct value for the conformal anomaly.

\kevin{Given that we get the correct value for the conformal anomaly without gauge fixing, it must be that if we use this regularisation to calculate in the standard way by gauge fixing, the resulting extra contributions all cancel amongst themselves. Adapting the earlier development of such a gauge fixed framework \cite{Arnone:2000bv, Arnone:2000qd,Arnone:2001iy}, we verify that this is the case. The cancellations arise because the gauge fixing must be done for the full local supergroup, which thus also implements Parisi-Sourlas supersymmetry.}


\section{Differential operators}
\label{sec:differentialOps}

Before proceeding, it is useful to collect together properties of the curved space differential operators that will naturally appear when working with gauge fields. For scalar fields $\omega$, the operator that naturally appears, \eg as the kernel in the kinetic term, is just the Laplace-Beltrami operator
\be 
\label{Delta0}
\Delta_0\, \omega := - \nabla^\mu\nabla_\mu\, \omega.
\ee
(With the sign, it is positive semi-definite.)
However for a, \eg $U(1)$, gauge field\footnote{For simplicity, in this section we write $A_\mu\equiv A^1_\mu$, and trust the reader not to be confused with later usage.} the kernel from
the simplest action
\be 
\label{actionU1}
\frac14\!\int\!\!d^Dx \sqrt{g}\, F_{\mu\nu} F^{\mu\nu} = \frac12\!\int\!\!d^Dx \sqrt{g}\, A^{\mu} \Delta_1^T A_\mu,
\ee
(where $F=dA$, or in components $F_{\mu\nu} =\nabla_\mu A_\nu -\nabla_\nu A_\mu$), is the differential operator $\Delta^T_1=\delta d$, where $d$ is the exterior derivative, and $\delta$ the co-differential. In components:
\beq
\label{Delta1T}
\Delta^T_1 A_{\mu} := \Delta_1 A_{\mu} + \nabla_{\mu} \nabla^{\nu} A_{\nu} = -\nabla^2 A_{\mu} + \nabla^{\nu} \nabla_{\mu} A_{\nu}.
\eeq
Here we have also introduced the (positive semi-definite) Laplace--de Rham operator 
\be (d+\delta)^2=d\delta+\delta d, \ee 
which on a one-form is explicitly
\beq
\label{Delta1}
 \Delta_1 A_{\mu} := -\nabla^2 A_{\mu} + R_{\mu}\,^{\nu} A_{\nu}
\eeq
(coinciding with the Lichnerowicz Laplacian). 
Abelian gauge invariance (\ie $d^2=0$) ensures that $\Delta^T_1$ annihilates longitudinal one-forms, as is easily explicitly verified:
\beq
\Delta^T_1 \nabla_{\mu} \omega =  -\nabla^2 \nabla_{\mu} \omega + \nabla^{\nu} \nabla_{\mu}\nabla_{\nu} \omega = 0.
\eeq
On the other hand since $d$ and $\delta$ commute with $(d+\delta)^2$, while on a scalar field de Rham $=$ Beltrami:
\be
(d+\delta)^2\omega = \delta d\,\omega = \Delta_0\,\omega,
\ee
we must have:
\beq
\label{Delta-properties}
\nabla^\mu \Delta_1 A_\mu =  \Delta_0  \nabla^\mu  A_\mu,\qquad
\Delta_1 \nabla_\mu\, \omega = \nabla_\mu \Delta_0\, \omega,
\eeq
as is also readily verified using the component formulae. Thus using \eqref{Delta1T}, we see that $\Delta_1$ and $\Delta^T_1$ commute.
Ignoring normalisable zero-modes (or working on a manifold which has none), $\Pi_L = d \frac1{(d+\delta)^2} \delta$ is a longitudinal projector for one-forms, equivalently
\beq
\Pi_{L} A_{\mu}:= - \nabla_{\mu} \frac{1}{\Delta_0}  \nabla^{\nu} A_{\nu}.
\eeq
Therefore the transverse projector is
\beq
\Pi_{T} := 1 - \Pi_{L}.
\eeq
By $d$, $\delta$ algebra, or using \eqref{Delta-properties}, we have
\beq
\Delta_1 \Pi_{T} A_{\mu} = \Pi_{T} \Delta_1 A_\mu = \Delta_1^T A_\mu,
\ee
%
\ie $\Delta_1^T$ is just the transverse projection of $\Delta_1$.
Splitting 
\be 
A_{\mu} = \Pi_T A_\mu + \Pi_L A_\mu =: A_{\mu}^T + \nabla_{\mu} A^L,
\ee
the transverse eigenmodes of $\Delta_1$ are the non-zero eigenmodes  of $\Delta_1^T$, while longitudinal eigenmodes of $\Delta_1$ are eigenmodes
of  $\Delta_0 A^L = \lambda A^L$, since then
\beq
\Delta_1 \nabla_{\mu} A^L = \lambda  \nabla_{\mu} A^L.
\eeq
As a result a trace involving $\Delta_1$ projected into the transverse modes can be expressed as
\beq
\label{trace-T}
\Tr_{T} f(\Delta_1) \equiv \Tr\, \Pi_T f(\Delta_1) = \Tr f(\Delta_1) - \Tr f(\Delta_0),
\eeq 
while the trace over the longitudinal sector is
\beq
\label{trace-L}
\Tr_{L} f(\Delta_1) \equiv  \Tr f(\Delta_1) - \Tr \Pi_T f(\Delta_1)  = \Tr f(\Delta_0).
\eeq 

\section{Manifestly gauge invariant flow equation 
on a curved spacetime}
\label{sec:flow}

We give a brief review of manifestly gauge invariant flow equations for Yang-Mills theory, making some minimal adaptations so that they apply to Maxwell theory propagating in a curved spacetime. As explained in the introduction, this will actually yield a $U(1)\times U(1)$ theory, where the second copy has wrong sign action. From this we can nevertheless extract the one-loop pure gravitational contribution. Then in sec. \ref{sec:improved} we will give an improved flow equation which leaves behind only a single physical Maxwell gauge field.

Recall that the basic idea is that the flow of the Boltzmann measure $\exp(-S)$ should be a 
total functional derivative, \ie for some generic fields $\phi$:
\be
\label{reparam}
\Lambda\partial_\Lambda \,{\rm e}^{-S} = {\delta\over\delta\phi}\left(\Psi
\,{\rm e}^{-S}\right)
\ee
(corresponding to the statement that each RG step is equivalent to an
infinitesimal field redefinition $\phi\mapsto \phi+\Psi\,
\Lambda^{-1} \delta\Lambda$) \cite{Morris:1998kz,Latorre:2000qc}. Importantly, this ensures that the
partition function ${\cal Z}=\int\!\!{\cal D}\phi\, \exp(-S)$, and
hence the physics derived from it, is invariant under the RG flow. Working with a fixed background metric $g_{\mu\nu}$ and in general $D$ dimensions, we will show how to generalise the previous formulations for Yang-Mills while preserving this crucial feature, and in such a way that the solution remains straightforwardly calculable. 

As we sketched in the introduction the previous formulations were developed over a number of years to cope with the most general cases. For our purposes we can closely follow the one set out in ref. \cite{Arnone:2002cs}. Indeed we will see that the flow equation still takes the generic form
\be
\label{sunnfl}
\ldl S  =
- a_0[S,\Sigma_\g]+a_1[\Sigma_\g],
\ee
where 
\be
\label{Sigma}
\Sigma_\g=\g^2S-2\hS,
\ee 
$\hS$ being the so-called seed action, and $\g$ being the gauge coupling which, since we work in general $D$ spacetime dimensions, has mass dimension $2-D/2$. The coupling has been factored out so that it plays the r\^ole of a loop counting parameter: the loop expansion of the effective action being given by 
\be
\label{Sloope}
S={1\over \g^2} S_0+S_1+\g^2 S_2 +\cdots,
\ee
where $S_0$ is the classical effective action, $S_1$ the one-loop
correction, and so on. This ensures that (super-)gauge invariance is manifestly maintained at each order in $\g$. As a consequence the super-covariant derivative is given by
\be
\label{gauge-covariant-derivative}
\mathcal{D}_\mu = \nabla_\mu -i\mathcal{A}_\mu,
\ee
The bilinear functional operator that generates the tree-level contributions is manifestly supergauge invariant:
\be 
\label{a0}
 a_0[S,\Sigma_\g] ={1\over2}\,\frac{\delta S}{\delta {\cal
A}^{\mu}}\{\dDelta^{\!\A\A}\}\frac{\delta \Sigma_\g}{\delta {\cal
A}_{\mu}}+{1\over2}\,\frac{\delta S}{\delta {\cal C}}\{\dDelta^{\C\C}\}
\frac{\delta \Sigma_\g}{\delta {\cal C}}, 
\ee
as is the linear functional that generates the loop corrections:
\be
\label{a1}
a_1[\Sigma_\g] = {1\over2}\,\frac{\delta }{\delta {\cal
A}^{\mu}}\{\dDelta^{\!\A\A}\}\frac{\delta \Sigma_\g}{\delta {\cal
A}_{\mu}} + {1\over2}\,\frac{\delta }{\delta {\cal C}}\{\dDelta^{\C\C}\}
\frac{\delta \Sigma_\g}{\delta {\cal C}}.
\ee
These expressions are exactly as in ref. \cite{Arnone:2002cs}.\footnote{\label{footnote:supergauge} Since super-gauge invariance ensures $\D_\mu \frac{\delta S}{\delta\A_\mu} = i[\C,\frac{\delta S}{\delta\C}]$, longitudinal terms can be absorbed into the $\C$ part \cite{Arnone:2002cs}.}
We now explain what the various terms mean. Unlike in ref. \cite{Arnone:2002cs}, we are interested in regularising $U(1)$ gauge theory. We therefore need to use a gauge field $\A$ valued as a generator of $U(1|1)$. The gauge field therefore takes the same form as given in eqn. \eqref{A}, except that here each field is thus a single component rather than a matrix and, unlike in the $SU(N|N)$ case, there is no resulting restriction to $\str \A=0$. The same goes anyway for the superscalar field
\be
\label{defC}
\C = \left( \begin{array}{cc} C^1 & D \\
			      \bar{D} & C^2       
	       \end{array}
       \right),
\ee
which will inflict spontaneous symmetry breaking on the supergauge invariance, breaking it down to the diagonal $U(1)\times U(1)$ carried by the bosonic gauge fields $A^i_\mu$, while supplying cutoff size masses to the complex fermionic $B_\mu$ field which thus turns into a gauge invariant Pauli-Villars regulator field. As explained in ref. \cite{Arnone:2002cs}, an elegant way to impose that the spontaneous symmetry breaking scale tracks the cutoff scale $\Lambda$, is to take $\C$ to be dimensionless, so we will do the same here. Then  we can choose a potential so that the effective vacuum expectation value can be \cite{Arnone:2002cs}:
\be 
\label{vev}
\langle \C \rangle = \sigma
\ee
at any scale $\Lambda$. Following previous convention, we write the third Pauli matrix as
\be
\label{sigma}
\sigma \equiv \sigma_3 = \begin{pmatrix} 1 & 0\cr 0 & -1 \end{pmatrix}.
\ee
This matrix appears frequently also as a result of the supergroup symmetry, for example through the supertrace:
\be 
\str X := \tr (\sigma X),
\ee
$X$ being a supermatrix and str being the supergroup invariant version the trace. The result of \eqref{vev} is precisely to give a mass $\sim\Lambda$ to the off-diagonal entries in $\A$ \ie to the complex $B$ field.

Since for the $U(1|1)$ theory, $\A$ is not subject to a constraint, both $\A$ and $\C$ functional derivatives are freely acting and are thus defined as follows:
\be
\label{dCdef}
{\delta \over {\delta\C}} := {
\left(\!{\begin{array}{cc} {\delta / {\delta C^1}} & - {\delta /
{\delta \bar{D}}} \\ {\delta / {\delta D}} & - {\delta
/ {\delta C^2}} \end{array}} \!\!\right)},
\ee
or in components 
\be
\label{Cdumbdef}
{\delta \over {\delta\C}}^i_{\gap j} :=
{\delta \over {\delta\C}^k_{\gap i}}\sigma^k_{\gap j},
\ee
the supergauge functional derivatives being defined in the same way. The advantage of this definition is that the $U(1|1)$ invariance remains manifest, for example we have:
\be
\label{sow}
{\partial\over\partial\C}\ \str \,\C Y = Y,
\ee
and thus 
\be 
\str X{\partial\over\partial\C}\ \str \,\C Y = \str XY,
\ee
and 
\be
\label{split}
\str {\partial\over\partial\C}X\C Y = \str X\ \str Y
\ee
where $X$ and $Y$ are arbitrary constant supermatrices \cite{Morris:2000fs,Arnone:2002cs}. 

 In order to maintain local supergauge invariance in \eqref{sunnfl} it is then only necessary to ensure that the bi-local kernels $\dDelta^{\!\A\A}(x,y)$ and $\dDelta^{\C\C}(x,y)$ 
are suitably covariantized by including $\A$ interactions, after which an invariant is constructed by taking an overall supertrace. This is essentially the meaning of the curly brackets. In fact it proved helpful to extend the definition so that\footnote{In non-abelian Yang-Mills, the couplings of $A^2$ and $A^1$ run differently, motivating further decorations \cite{Arnone:2005fb}, however this is not needed for the calculation pursued here.}
\be
\label{wev}
X\{W\}Y =  X\,\{W\}_{\!\!{}_\A}Y -{1\over4}
[\C,X]\,\{W_{m}\}_{\!\!{}_\A} [\C,Y],
\ee
where $X(x)$ and $Y(y)$ are supermatrix fields produced by the functional derivatives in \eqref{a0},  $W_m(x,y)$ is a new kernel that simplifies calculations in the broken phase, and $\{\cdots\}_{\!\!{}_\A}$ stands for the gauge covariantization just described. 
In flat space, the most general form of gauge covariantization is described in ref. \cite{Arnone:2002cs}, following \cite{Morris:2000fs,Morris:1999px,Arnone:2002qi}, and can be couched in terms of a path integral over Wilson lines. We will not need the details for this paper. However we will need a covariantization to cope with a non-trivial metric. The most general case can again be couched in terms of an integral over Wilson lines for the Levi-Civita connection, as remarked in ref. \cite{Morris:2016nda}.
In this latter paper we however made the simplest choice, promoting space-time partial derivatives to covariant derivatives in a prescribed way (corresponding implicitly to some specific choice of measure for the Wilson lines). We will do something similar in  this paper. Thus
 for  a scalar flat-space kernel:
\be
\label{kdef}
W(x,y) =\int\!\!{d^D\!p\over(2\pi)^D}
\,W(p^2,\Lambda)\,{\rm e}^{ip.(x-y)} \ =\ W(-\partial_x^2,\Lambda)\, \delta(x-y),
\ee
we make the replacement
\be 
\label{covariantize}
W(-\partial^2_x,\Lambda) \mapsto W(\Delta_{0\,x},\Lambda)/\sqrt{g(x)},
\ee
where $\Delta_0$ is the Laplace-Beltrami operator introduced in \eqref{Delta0}. Following the framework of \eg ref. \cite{Morris:2016nda}, the
factor of $1/\sqrt{g}$ is inserted to give the kernel the correct overall density of weight -1  so that combined with the $\sqrt{g}$ factors from the two functional derivatives in \eqref{a0} and after integrating over $x$ and $y$ (without further factors $\sqrt{g}$) the result is clearly generally covariant.  (Note that $\nabla_\alpha$ commutes with $1/\sqrt{g}$  which can have either $x$ or $y$ dependence, and thus the kernel is symmetric as assumed.)
Recognising that the vector kernel $\dDelta^{\!\A\A}$ is associated with one-forms, the elegant choice is to make this a function of the Laplace--de Rham operator $\Delta_1$, \cf \eqref{Delta1}. We will see in the remainder of the paper how this choice ensures that computations remain almost as simple as their flat-space counterparts. 
As in ref. \cite{Arnone:2002cs}, we discard the terms where the left-most $\C$ functional derivative in \eqref{a1} hits the $\C$ decorations in \eqref{wev}.
This can be imposed by a limiting procedure \cite{Morris:1998kz}, see also \cite{Morris:2000fs,Morris:1999px}. 

$\hS$ is used to determine the form of the classical effective kinetic
terms and the kernels $\dDelta$.  It therefore has to
incorporate the covariant higher derivative regularisation and allow
the spontaneous symmetry breaking we require. As we will review shortly, the kernels $\dDelta$ are determined 
by the requirement that after spontaneous
symmetry breaking, the two-point vertices of the classical effective
action $S_0$, and $\hS$ can be set equal.  This is imposed as a useful technical device,
since it allows classical vertices to be immediately solved in terms of already known quantities.  



\section{Kernels and two-point vertices in a curved background}

In this paper we are interested only in the one loop contribution to pure gravity. This arises by first solving for the classical action $S_0$. For this we extract the $1/\g^2$ part of \eqref{sunnfl}, using \eqref{Sigma} and \eqref{Sloope}:
\be 
\Lambda{\partial\over\partial\Lambda}S_0 = -a_0[S_0,S_0-2\hS].
\label{ergcl}
\ee
Then the one-loop piece $S_1$ can be solved for, by substituting back into the flow equation \eqref{sunnfl}. We see that the pure gravity contribution arises from $a_1[S_0-2\hS]$ where we need only the $\C$ and $\A$ two-point vertices in $S_0$ and $\hS$.  We also see we can dispense with the $U(1|1)$ gauge covariantization $\{\cdots\}_{\!\!{}_\A}$.

Just as in ref. \cite{Arnone:2002cs}, there are no $\A$ one-point vertices (\eg as a result of Poincar\'e invariance or charge conjugation invariance). Expanding around  $\C\mapsto \C+\sigma$, where by design $\C=\sigma$ is at the minimum of the potential, there are no  $\C$ one-point vertices either. Therefore we also do not need the $U(1|1)$ gauge covariantization in the classical flow \eqref{ergcl} of the two-point vertices. As just stated in the previous section, these are set equal to the seed action two-point vertices. Since $\hS$ is our choice, the flow actually serves to determine the kernels. Indeed specialising to the two-point vertices, \eqref{ergcl} now simply becomes
\be 
\Lambda{\partial\over\partial\Lambda}\hS = a_0[\hS,\hS]\qquad\hbox{(for two-point vertices)}.
\label{ergcl2}
\ee
Universal quantities are however independent of 
the choices made, which are part of the freedom in \eqref{reparam} to reparametrise the fields \cite{Morris:1998kz,Latorre:2000qc}. 

From \eqref{Sigma}, since $\g$ has mass dimension $2-D/2$, $\hS$ has dimension $4-D$, \ie the Lagrangian component has dimension four, independent of space-time dimension (similarly from \eqref{Sloope} for $S_0$). Since the above gives structurally the same equations for the flow of the two-point vertices as in ref. \cite{Arnone:2002cs}, we can therefore follow closely in this section the derivations given there.  We thus split the supermatrix fields into their block (off-)diagonal components $A_\mu=d_+\A_\mu$, $B_\mu= d_-\A_\mu$, $C=d_+\C$ and $D=d_-\C$ where
\be
\label{d}
{\rm d}_\pm X = {1\over2}(X\pm\sigma X\sigma).
\ee
Choosing a single supertrace form for $\hS$, we need to determine the differential operators that form the two-point vertices $\hS^{AA}_{\mu\nu}$, $\hS^{BB}_{\mu\nu}$, $\hS^{BD\sigma}_\mu$, $\hS^{CC}$ and $\hS^{DD}$ \cite{Arnone:2002cs}. In each case the superscript gives the order of the fields as they appear in the supertrace (up to cyclicity), for example $\hS^{BD\sigma}_\mu(\nabla)$ sits in the seed action as the term
\be 
\label{BD}
\str \,\,\sigma\!\! \int\!\!d^Dx \sqrt{g}\,   B^\mu \hS^{BD\sigma}_\mu D,
\ee
where we have used cyclicity of the supertrace to put $\sigma$ first. The flow equations resulting from \eqref{ergcl2} then take exactly the same form as in ref. \cite{Arnone:2002cs}:
\bea
\label{fl2}
\ldl \hS^{CC} &= &\hS^{CC}\dDelta^{CC}\hS^{CC},\nonumber\\
\ldl \hS^{AA} &= &\hS^{AA}\dDelta^{AA} \hS^{AA},\nonumber\\
\ldl\hS^{BB}_{\mu\nu} &= &\left(\hS^{BB}\dDelta^{BB}\hS^{BB}\right)_{\!\mu\nu}\!\!
+\hS^{BD\sigma}_\mu\dDelta^{DD}\hS^{BD\sigma}_\nu,\nonumber\\
\ldl\hS^{BD\sigma}_\mu &= &\hS^{BB}\dDelta^{BB}\hS^{BD\sigma}_\mu
+\hS^{BD\sigma}_\mu\dDelta^{DD}\hS^{DD},\nonumber\\
\ldl\hS^{DD} &= &\hS^{DB\sigma\, \mu}\dDelta^{BB}\hS^{BD\sigma}_\mu
+\hS^{DD}\dDelta^{DD}\hS^{DD},
\eea
(where the last two follow from the third by spontaneously broken supergauge invariance) and where the kernels
\be
\label{bzptA}
\dDelta^{AA} = \dDelta^{\A\A},\quad
\dDelta^{CC} = \dDelta^{\C\C}
\ee
and
\be 
\label{bzptB}
\dDelta^{BB} = \dDelta^{\A\A}+\dDelta^{\A\A}_m,\quad
\dDelta^{DD} = \dDelta^{\C\C}+\dDelta^{\C\C}_m,
\ee
are also of the same form
except that following the standard convention in gravitation, see \eg eqns. \eqref{actionU1} and \eqref{Delta1T},
 the indices on the differential operators $\hS^{AA}$, $\hS^{BB}$, $\dDelta^{\A\A}$, and $\dDelta^{BB}$ have been suppressed where it is unambiguous to do so, and in the last line of \eqref{fl2} we recognise that the first vertex that appears
on the right hand side needs now to be distinguished from the second (as discussed below).

The only changes to the solutions found in ref. \cite{Arnone:2002cs} are thus induced by the covariantizations of the seed-action two-point vertices required to cope with the background metric $g_{\mu\nu}$. We are free to choose these. Thus we set
\be 
\label{hSAC}
\hS^{AA} =  2 \Delta^T_1/c_1, \quad \hS^{CC} = \Lambda^2\Delta_0/\ct_0 + 2\lambda\Lambda^4,
\ee
and 
\be 
\label{hSBD1}
\hS^{BB} =  2 \Delta^T_1/c_1+4\Lambda^2/\ct_1,\quad
\hS^{DD} = \Lambda^2\Delta_0/\ct_0,
\ee
where $\lambda>0$ is a constant dimensionless parameter \cite{Arnone:2002cs}, and $c_i=c(\Delta_i/\Lambda^2)$ and $\ct_i = \ct(\Delta_i/\Lambda^2)$ are cutoff functions \cite{Arnone:2002cs} of the appropriate Laplace--de Rham operator. Similarly
\be
\label{hSBD}
\hS^{BD\sigma}_\mu = 2i\Lambda^2 \nabla_\mu \,\ct_0^{-1} = 2i\Lambda^2 \ct_1^{-1} \,\nabla_\mu\qquad\mathrm{and}\qquad \hS^{DB\sigma}_\mu = 2i\Lambda^2 \nabla_\mu \,\ct_1^{-1} = 2i\Lambda^2 \ct_0^{-1} \,\nabla_\mu,
\ee
where we used \eqref{Delta-properties}. The second version follows from integration by parts in \eqref{BD}, followed by cycling the supertrace and anticommuting $\sigma$. In ref. \cite{Arnone:2002cs} it was not needed, since in flat space the two coincide.  

Substituting \eqref{hSAC}, \eqref{hSBD1} and \eqref{hSBD} into \eqref{fl2}, yields the kernels, and thus the integrated kernels defined via
\be 
\label{prop}
\dDelta = -\ldl \prop.
\ee
The integration constant is determined by ensuring that the corresponding $\prop$ vanish for large eigenvalue. We now show that we get for the (integrated) kernels, the obvious covariantization of the results found in \cite{Arnone:2002cs}. The first two equations in \eqref{fl2} are solved by straightforward integration:
\be 
\label{propCCAA}
\prop^{CC} = \left(\hS^{CC}\right)^{-1}=\frac1{\Lambda^2} \frac{\ct_0}{\Delta_0+2\lambda\Lambda^2\ct_0},\qquad \prop^{AA} = \frac{c_1}{2\Delta_1},
\ee
where the second is the inverse of $\hS^{AA}$ in the transverse space. Multiplying the third equation in \eqref{fl2} by the transverse projector $\Pi_T$, isolates
\be 
\label{propBB}
\prop^{BB} \equiv \prop^{BB}(\Delta_1) = \frac12\frac{c_1\ct_1}{\Delta_1\ct_1+2\Lambda^2c_1},
\ee
the inverse of $\Pi_T\hS^{BB}$ in the transverse space. Substituting for the vertices and rearranging the last equation in \eqref{fl2} gives
\be 
\label{propDD}
\prop^{DD} = {\ct_0}/({\Lambda^2\Delta_0})-4\prop^{BB}_0/\Delta_0 = \frac1{\Lambda^2}\frac{\ct^2_0}{\ct_0\Delta_0+2\Lambda^2c_0},
\ee
where $\prop^{BB}_0 \equiv \prop^{BB}(\Delta_0)$, \ie \eqref{propBB} with $\Delta_1$ replaced by $\Delta_0$. The above formulae are indeed direct maps of the results in ref. \cite{Arnone:2002cs}.


\section{Twice the manifestly gauge invariant conformal anomaly}
\label{sec:one-loop}

From \eqref{sunnfl}, \eqref{Sigma} and \eqref{Sloope}, and the equality of classical and seed-action two-point vertices, we have that the purely gravitational part of the one-loop effective action is computed from
\be 
\label{flone-loop}
\ldl S_1 = -a_1[\hS].
\ee
The functional derivatives in \eqref{a1} are evaluated using \eqref{split}, after expressing the block (off)-diagonal fields in terms of the originals via \eqref{d}. The supergroup contribution is then $\pm\half(\str\sigma)^2=\pm 2$, \ie the signed number of each flavour. Combined with the sign in \eqref{flone-loop} and the $1/2$ from \eqref{a1}, we thus have that the purely gravitational piece satisfies:
\be 
\label{dS1}
\ldl S_1 = \Tr\left[ -\hS^{AA}\dDelta^{AA} +\hS^{BB}\dDelta^{BB} -\hS^{CC}\dDelta^{CC} +\hS^{DD}\dDelta^{DD} \right].
\ee
where $\Tr$ stands for a space-time trace, taking into account the relevant Lorentz representation. Thus for $A$ this is a trace over transverse modes, for $B$ a trace over all vector modes, and for $C$ and $D$ it is a trace over scalar modes. The bosonic contributions are straight-forward to simplify using the equations of the previous section:
\be 
\Tr\left[ \hS^{AA}\dDelta^{AA}+\hS^{CC}\dDelta^{CC}\right] = \ldl\Tr\ln(\Delta^T_1/c_1) +\ldl\Tr\ln(\Lambda^2\Delta_0/\ct_0 + 2\lambda\Lambda^4).
\ee
Noting the first equation in \eqref{propDD}, we have
\be 
\Tr\, \hS^{DD}\dDelta^{DD} = \ldl \Tr \ln(\Lambda^2\Delta_0/\ct_0) -4\Lambda^2\,\Tr\,\dDelta^{BB}_0/\ct_0,
\ee
while using the first equation of \eqref{Delta1T} and cyclicity of the spacetime trace,
\be 
\Tr\, \hS^{BB}\dDelta^{BB} = \ldl \Tr\ln(\Delta_1/c_1+2\Lambda^2/\ct_1)-2\Tr\,\Delta_0\dDelta^{BB}_0/c_0.
\ee
Combining the last terms in the above two equations gives $\hS^{BB}_0\dDelta^{BB}_0$, which again simplifies.
Thus, substituting everything back into \eqref{dS1}, we can trivially integrate with respect to $\Lambda$.
Also cancelling $\Tr\ln\Lambda^2$ between $D$ and $C$ sectors, we thus get
\be
S_1 =  \Tr\left[ -\ln(\Delta^T_1/c_1) +\ln(\Delta_1/c_1+2\Lambda^2/\ct_1) -\ln(\Delta_0/c_0+2\Lambda^2/\ct_0) +\ln(\Delta_0/\ct_0) -\ln(\Delta_0/\ct_0 + 2\lambda\Lambda^2) \right].
\ee
From \eqref{trace-L}, the third term on the right hand side is just subtracting the longitudinal $B$ contribution. Indeed using \eqref{trace-T}, we can alternatively write:
\be 
\label{twiceStandard}
S_1 = -\Tr_T\left[ \ln(\Delta^T_1/c_1) -\ln(\Delta_1^T/c_1+2\Lambda^2/\ct_1) \right] +\Tr\left[\ln(\Delta_0/\ct_0) -\ln(\Delta_0/\ct_0 + 2\lambda\Lambda^2) \right].
\ee
In this form we recognise that the result coincides with twice what would be produced by more conventional calculational methods, reflecting the fact that we have two copies $A^i$ of the $U(1)$ gauge field. Thus the first trace is the contribution of two transverse vector fields regularised by covariant higher derivatives and Pauli-Villars. The second trace coincides with twice the Jacobian from the change of variables $A_\mu = A^T_\mu +\nabla_\mu\omega$, again regulated by covariant higher derivatives and a Pauli-Villars field. 

Using \eqref{trace-T} we can map \eqref{twiceStandard} to a calculation which is even closer to a standard textbook exposition. By replacing the transverse trace by a trace over the full vector representation we get:
\besp 
\label{twiceStandard}
S_1 = -\Tr\left[ \ln(\Delta_1/c_1) -\ln(\Delta_1/c_1+2\Lambda^2/\ct_1) \right]\\ 
+\Tr\left[ \ln(\Delta_0/c_0) -\ln(\Delta_0/c_0+2\Lambda^2/\ct_0) \right]
+\Tr\left[\ln(\Delta_0/\ct_0) -\ln(\Delta_0/\ct_0 + 2\lambda\Lambda^2) \right].
\eesp
Now from \eqref{Delta1} the first term is the one loop contribution for two $U(1)$ gauge fields in Feynman gauge, regulated by covariant higher derivatives and a Pauli-Villars field, whereas the second two terms can be identified with the ghost contributions regulated by covariant higher derivatives and Pauli-Villars fields (with the option to choose different parameters for the regularisation in the second ghost action).

Either way we see that, following now standard treatments, for example computing the Schwinger-Dewitt coefficients in a heat kernel expansion (see \eg \cite{Duff1994}), will yield twice the trace anomaly contribution from massless vector fields: 
\be 
\label{text-book-x2}
(4\pi)^2 g^{\alpha\beta}\langle T_{\alpha\beta}\rangle = 2b \left(C^{\mu\nu\rho\sigma}C_{\mu\nu\rho\sigma}+\frac23\nabla^2 R\right) + 2b'\, {}^*\! R_{\mu\nu\rho\sigma} {}^*\! R^{\mu\nu\rho\sigma}\,,
\ee
where $C$ is the Weyl tensor, and $b=1/10$ and $b'=-31/180$ the standard values.

\section{Twice the contribution to the gravitational beta functions}
\label{sec:beta}

The coefficients $b$ and $b'$ also appear in the gravitational beta functions induced by the gauge fields. These are obtained by taking a derivative with respect to $\Lambda$ of \eqref{twiceStandard} obtaining the traces
\beq \label{dS1W}
\Lambda \partial_{\Lambda} S_1 =    \Tr_1 [ W_0(\Delta_0)]  +\Tr_0 [ W_0(\Delta_0)] 
\eeq
with the functions
\bea \label{Wfunctions}
  W_1(\Delta_1)
     &=&  \frac{4 \left(\tilde{c}\left(\frac{\Delta _1}{\Lambda ^2}\right) \left(\Lambda
   ^2 c\left(\frac{\Delta _1}{\Lambda ^2}\right)-\Delta _1 c'\left(\frac{\Delta
   _1}{\Lambda ^2}\right)\right)+\Delta _1 c\left(\frac{\Delta _1}{\Lambda
   ^2}\right) \tilde{c}'\left(\frac{\Delta _1}{\Lambda
   ^2}\right)\right)}{\tilde{c}\left(\frac{\Delta _1}{\Lambda ^2}\right)
   \left(\Delta _1 \tilde{c}\left(\frac{\Delta _1}{\Lambda ^2}\right)+2 \Lambda
   ^2 c\left(\frac{\Delta _1}{\Lambda ^2}\right)\right)} \\
W_0(\Delta_0) &=& \frac{4 \Delta _0 \tilde{c}\left(\frac{\Delta _0}{\Lambda ^2}\right)
   c'\left(\frac{\Delta _0}{\Lambda ^2}\right)-4 c\left(\frac{\Delta _0}{\Lambda
   ^2}\right) \left(\Delta _0 \tilde{c}'\left(\frac{\Delta _0}{\Lambda
   ^2}\right)+\Lambda ^2 \tilde{c}\left(\frac{\Delta _0}{\Lambda
   ^2}\right)\right)}{\tilde{c}\left(\frac{\Delta _0}{\Lambda ^2}\right)
   \left(\Delta _0 \tilde{c}\left(\frac{\Delta _0}{\Lambda ^2}\right)+2 \Lambda
   ^2 c\left(\frac{\Delta _0}{\Lambda ^2}\right)\right)} \nonumber \\
   && + \frac{4 \lambda 
   \left(\Delta _0 \tilde{c}'\left(\frac{\Delta _0}{\Lambda ^2}\right)-\Lambda
   ^2 \tilde{c}\left(\frac{\Delta _0}{\Lambda ^2}\right)\right)}{2 \lambda 
   \Lambda ^2 \tilde{c}\left(\frac{\Delta _0}{\Lambda ^2}\right)+\Delta _0}\,. 
  \eea
 Evaluating the traces \eqref{dS1W} using the early time heat kernel expansion up to second order in curvature we have
\bea
\Tr [W_0(\Delta_0)] &=&  \frac{1}{(4\pi)^2} \left(  Q_2[W_1]  B_{0}(\Delta_1) + Q_1[W_1] B_{1}(\Delta_1)    +   Q_0[W_0]   B_{2}(\Delta_0) + ...  \right)\,,
\eea
and
\bea
\Tr [W_1(\Delta_1)] &=&  \frac{1}{(4\pi)^2} \left( Q_2[W_1]  B_{0}(\Delta_1) + Q_1[W_1] B_{1}(\Delta_1)   +   Q_0[W_0] B_{2}(\Delta_1) \right) + ... \,,
\eea
where $B_n(\Delta_i)$ are the traced heat kerne coefficients for the operators $\Delta_0$ and $\Delta_1$ and $Q_m [W_i]$ are functionals of the the functions \eq{Wfunctions}.
Explicitly the heat kernel coefficients are given by 
\bea
B_0(\Delta_1) &=&4 \int d^Dx \sqrt{g} \,, \nonumber \\
B_1(\Delta_1) &=&   -\frac{1}{3}  \int d^Dx \sqrt{g}   R \,, \nonumber \\
B_2(\Delta_1) &=&    \int d^4x \sqrt{g}   \left(  -\frac{1}{30} \nabla^2 R  + \frac{7}{60} C^2 - \frac{8}{45}  {}^*\! R_{\mu\nu\rho\sigma} {}^*\! R^{\mu\nu\rho\sigma} + \frac{1}{36} R^2  \right) \,, \nonumber \\
B_0(\Delta_0) &=&  \int d^Dx \sqrt{g} \,,  \nonumber \\
B_1(\Delta_0) &=& \frac{1}{6}  \int d^Dx \sqrt{g}  R \,, \nonumber \\
B_2(\Delta_0) &=&    \int d^4x \sqrt{g}   \left(   \frac{1}{180}  \left( \frac{3}{2} C_{\mu\nu\rho\sigma} C^{\mu\nu\rho\sigma}  -\frac{1}{2}  {}^*\! R_{\mu\nu\rho\sigma} {}^*\! R^{\mu\nu\rho\sigma}  \right) + \frac{1}{72} R^2  + \frac{1}{30} \nabla^2 R  \right) \nonumber \,.
\eea
 For $m>0$ the $Q_m[W_i]$ functionals are given by the scheme dependent integrals
\beq
Q_m[W_i] =  \frac{1}{\Gamma(m)} \int_0^\infty dz z^{m-1} W_i(z)  
\eeq
whereas the $Q_0$ functionals are given by 
\bea
Q_0[W_1] = W_1(0) = 2\,,  \,\,\,\,\,\,\,  Q_0[W_0] = W_0(0) = -4
\eea
which are independent of the choice of cutoffs $c$ and $\tilde{c}$. Consequently we find that the logarithmic  terms give the trace anomaly 
\be
\label{beta}
\Lambda \partial_{\Lambda} S_1 =  2  \frac{1}{(4\pi)^2}  \int d^4x \sqrt{g} \left[  \Lambda^4 a_0  + a_1 \Lambda^2 R +  b \left(C^{\mu\nu\rho\sigma}C_{\mu\nu\rho\sigma}+  \nabla^2 R\right) + b'\, {}^*\! R_{\mu\nu\rho\sigma} {}^*\! R^{\mu\nu\rho\sigma} +... \right]
\ee
with  $b=1/10$ and $b'=-31/180$ as in \eqref{text-book-x2}. 
The scheme dependent coefficients are given by $a_0  = 4 Q_2[W_1]  + Q_2[W_0]$ and $a_1 =  - \frac{1}{3} Q_1[W_1]  + \frac{1}{6} Q_1[W_0] $ are non-universal since they depend on the form of the cutoff functions $c$ and $\tilde{c}$ and determine the running of the vacuum energy and the Newton's constant.

\section{Spontaneous symmetry breaking by the vector representation}
\label{sec:SSB}

We have just seen that we get twice the desired gravitational contribution, because the $A^2$ part of the regularisation structure remains massless. We now repair this problem with the regularisation.  

Up until now we have treated the two diagonal entries in \eqref{A} equally, using $A = d_+\A$, where $d_+$ is defined in \eqref{d}. Splitting this further down to $A= A^1\sigma_+ +A^2\sigma_-$, where $\sigma_\pm = (\one\pm\sigma)/2$, and
\be 
\label{defA1A2}
A^1 = \sigma_+\A\sigma_+,\qquad A^2 = \sigma_-\A\sigma_-,
\ee
we want to give a mass to $A^2$ while leaving $A^1$ massless. Therefore we must spontaneously break the $\sigma_-$ direction while leaving $\sigma_+$ direction unbroken. That is not possible using only commutators (roughly speaking, the supergroup adjoint representation) since $\one$ commutes with anything and thus
\[
[\sigma_+,X] =-[\sigma_-,X]\,,
\]
for an arbitrary supermatrix. The next simplest thing to do therefore is to introduce a `fundamental' \aka vector representation,\footnote{Since $SU(N|N)$ is an example of supergroup that is reducible but not decomposable, terms such as fundamental and adjoint carry caveats \cite{Arnone:2001iy}.} redefining
\be 
\label{new-C}
\C = \begin{pmatrix}
D \\ C
\end{pmatrix}\,.
\ee
For regularising $SU(N|N)$ this would not have worked, firstly because the number of degrees of freedom are incorrect to be eaten by $B$ (and $A^2$), and secondly again because $\one$ commutes with anything. We pause briefly to sketch why the latter property would have led to an issue. In $SU(N|N)$, the supergauge field can be alternatively expanded as $\A=\A^0\one+\A^AT_A$, where the generators $T_A$ are both traceless and supertraceless since $\str \A=0$ has forbidden the appearance of a $\sigma$ term. When interactions are built on commutators, this furthermore implies that $\A^0$ appears nowhere in the action, resulting in the ``no-$\A^0$'' shift symmetry $\delta\A^0_\mu(x)=\lambda_\mu(x)$ \cite{Arnone:2001iy,Arnone:2002cs} which then needs to be imposed as a consistency condition. (The alternative procedure of redefining the Lie bracket in the gauge sector to exclude terms $\propto\one$ leads to equivalent consistency conditions \cite{Arnone:2001iy,Arnone:2002cs}.) The second problem with breaking the $SU(N|N)$ symmetry using a representation \eqref{new-C} is that $\A^0$ will now couple to the action exclusively through such terms.  It would therefore work as a Lagrange multiplier field and force an unpromising non-linear constraint. This issue is analogous the problems which arise in regularised $SU(N|N)$ theory, if one attempts to impose that the matrix scalar field \eqref{defC} is supertraceless \cite{Arnone:2001iy}.

But $SU(N|N)$ is not what we are interested in here. Instead it turns out that the single superfield representation \eqref{new-C} is exactly what is needed. First we notice that the number of degrees of freedom is just right to give $B_\mu$, $\bar{B}_\mu$ and $A^2_\mu$ masses, if $\C$ is taken to be complex, and if the fermionic directions are broken and one of the bosonic directions is broken. In particular, to achieve the breaking of $A^2$'s $U(1)$, we need to get a vacuum expectation value in the bottom half. That is why we placed the fermionic component in the upper half and the bosonic component in the lower half. Now suppose that
\be 
\label{vev-again}
\langle\C\rangle = \begin{pmatrix}
0 \\ 1
\end{pmatrix}\,.
\ee
Under the supergroup, the fields transform as: $\delta\A_\mu = [\D_\mu,\Omega]$ and $\delta\C = i\Omega\,\C$. Writing \cite{Arnone:2002cs}
\[
\Omega = \begin{pmatrix}
\omega^1 & \tau \\
\bar{\tau} & \omega^2
\end{pmatrix}
\]
(where the $\omega^i$ are real and $\tau$ is complex), we see that the Goldstone modes are
\[
i\Omega \begin{pmatrix}
0 \\ 1
\end{pmatrix} = i \begin{pmatrix}
\tau \\ \omega^2
\end{pmatrix}\,,
\]
so indeed the fermionic and $A^2$ directions are completely broken as required. Furthermore, shifting $\C\mapsto \langle\C\rangle+\C$, we see that unitary gauge thus consists in setting $\C=\langle\C\rangle (1+C_R/\sqrt{2})$, where $C_R/\sqrt{2}$ is the real part of $C$. Since from \eqref{gauge-covariant-derivative},
\[
\D_\mu\bar{\C} = \nabla_\mu\bar{\C}+i\bar{\C}\A_\mu\,,
\]
we have that in unitary gauge the kinetic term for $\C$ reduces to
\be 
\label{kinC}
-\Lambda^2\D_\mu\bar{\C}\D^\mu\C = -\frac{\Lambda^2}2 \nabla_\mu C_R\,\nabla^\mu C_R+
\Lambda^2(1+C_R/\sqrt{2})^2 \left( B^\mu\bar{B}_\mu -A^2_\mu A^{2\,\mu}\right)\,.
\ee
The choice of sign for the kinetic term, thus provides the right sign mass for both $B$ and $A^2$, and shows that $C_R$ is a regulator field with the wrong-sign action. (Expanding the supertrace one sees that $\nabla B\nabla\bar{B}$ is the order that appears in the kinetic term with positive sign.\footnote{\label{footnote:signid}The choice of sign in \eqref{kinC} is already implicit in the previous formulation, as can be seen by taking the right hand column of \eqref{defC} and forming the supertrace.}) As before \cite{Arnone:2002cs}, we can ensure this Higgs field gets a mass term that tracks the cutoff, by making it dimensionless and assuming an appropriate potential. The minimal Lagrangian would be
\be 
\label{C-minimal}
\cL_\C = -\Lambda^2\D_\mu\bar{\C}\D^\mu\C -\frac{\lambda}{4}\Lambda^4 (\bar{\C}\C-1)^2\,,
\ee
supplying a mass ($\lambda\Lambda^2$) for the Higgs field $C_R$ in the broken phase.

\section{Manifestly gauge invariant flow equation for Maxwell theory}
\label{sec:improved}

Now we implement this spontaneous symmetry breaking scheme within a manifestly gauge invariant flow equation. Following sec. \ref{sec:flow} we keep the definitions \eqref{sunnfl}, \eqref{Sigma} and \eqref{Sloope} but rather evidently \eqref{a0} and thus \eqref{a1} should be replaced by
\bea 
\label{a0-again}
 a_0[S,\Sigma_\g]  &=& {1\over2}\,\frac{\delta S}{\delta {\cal
A}^{\mu}}\{\dDelta^{\!\A\A}\}\frac{\delta \Sigma_\g}{\delta {\cal
A}_{\mu}}+{1\over2}\,\frac{\delta S}{\delta {\cal C}}\{\dDelta^{\C\C}\}
\frac{\delta \Sigma_\g}{\delta \bar{{\cal C}}}, \\
\label{a1-again}
a_1[\Sigma_\g] &=& {1\over2}\,\frac{\delta }{\delta {\cal
A}^{\mu}}\{\dDelta^{\!\A\A}\}\frac{\delta \Sigma_\g}{\delta {\cal
A}_{\mu}} + {1\over2}\,\frac{\delta }{\delta {\cal C}}\{\dDelta^{\C\C}\}
\frac{\delta \Sigma_\g}{\delta \bar{{\cal C}}},
\eea
where the functional derivatives with respect to $\C$ and $\bar{\C}$ are just the functional derivatives with respect to the components except that the functional derivative in $\delta S/\delta\C := \delta_r S/\delta\C$ should be regarded as acting on the action (and thus also $\Sigma_\g$) from the right, so as not to introduce unnecessary signs into the Grassmann components.\footnote{Equivalently one takes a left-derivative using the bottom row of \eqref{dCdef}, and includes an overall minus sign for the overall supertrace, consistent with the identification in footnote \ref{footnote:signid}.}
The covariantization of the kernels replaces \eqref{wev} with 
\be
\label{wev-A}
\frac{\delta}{\delta\A^\mu}\{\dDelta^{\A\A}\}\frac{\delta}{\delta\A_\mu} =  \frac{\delta}{\delta\A^\mu}\,\{\dDelta^{\A\A}\}_{\!\!{}_\A}\frac{\delta}{\delta\A_\mu} -\bar{\C}\frac{\delta}{\delta\A^\mu}\,\{\dDelta^{\A\A}_{m}\}_{\!\!{}_\A}\frac{\delta}{\delta\A_\mu}\C\,,
\ee
where $W_m$ therefore now propagates a vector representation, and
\be
\label{wev-C}
\frac{\delta}{\delta\C}\{\dDelta^{\C\C}\}\frac{\delta}{\delta\bar{\C}} =  \frac{\delta}{\delta\C}\,\{\dDelta^{\C\C}\}_{\!\!{}_\A}\frac{\delta}{\delta\bar{\C}} +\left(\frac{\delta}{\delta\C}\otimes\C -\bar{\C}\otimes \frac{\delta}{\delta\bar{\C}}\right)\!\{\dDelta^{\C\C}_{m}\}\!\left(\frac{\delta}{\delta\C}\otimes\C -\bar{\C}\otimes \frac{\delta}{\delta\bar{\C}}\right)\,,
\ee
where $\dDelta^{\C\C}_m$ thus propagates the matrix ($\sim$ `adjoint') representation. This tensor-product type $\C$ decoration can be understood as arising from supergauge invariance (compare footnote \ref{footnote:supergauge}):
\be 
\str\left(\!\Omega\, \D_\mu \frac{\delta S}{\delta\A_\mu}\right) = i\left( \frac{\delta S}{\delta\C}\Omega\,\C - \bar{\C}\Omega\frac{\delta S}{\delta\bar{\C}}\right)\,,
\ee
\ie as before the $\C$ decoration in \eqref{wev-C} can be exchanged for longitudinal terms in \eqref{wev-A}.


The rest of the definition of the flow equation is as in sec. \ref{sec:flow}. In particular, functional derivatives do not act on the terms that decorate the kernels, only on the relevant action $S$ or $\Sigma_\g$. Clearly the resulting flow equation manifestly preserves local $U(1|1)$ invariance.

\section{Kernels and two-point vertices for the Maxwell flow equation}
\label{sec:kernels-again}

At the two-point level in the broken phase, the $\C$ and $\bar{\C}$ decorations are replaced with the vacuum expectation value \eqref{vev-again}. Defining $C=(C_R+iC_I)/\sqrt2$ and $D$ via the components in \eqref{new-C}, and the vector fields by their components, or equivalently and more conveniently via $B=d_-\A$ and \eqref{defA1A2}, we can find the resulting two-point flow equations analogous to \eqref{fl2}. Bearing in mind that we ensure that \eqref{new-C} solves the effective equations of motion, unpacking \eqref{wev-A} and \eqref{wev-C} reveals that the $B$ and $D$ kernels again collect as in \eqref{bzptB}, the $A_2$ and $C_I$ kernels coincide with these, and the new unbroken sector has adopted the old $A$ and $C$ expressions:
\be
\label{bzpt1}
\dDelta^{A_1A_1} = \dDelta^{\A\A},\quad 
\dDelta^{C_RC_R} = \dDelta^{\C\C},\quad
\dDelta^{A_2A_2}= \dDelta^{BB},\quad 
\dDelta^{C_IC_I} = \dDelta^{DD}.
\ee
As before, the seed action is our choice (subject to preservation of all the required symmetries, in particular spontaneously broken $U(1|1)$ invariance), and requiring that the two-point vertices of the classical effective action and the seed action can be set equal, then determines the kernels. The $\hS$ kinetic terms for the vector fields follow from covariant higher derivative regularisation of the super-field strength squared \cite{Arnone:2002cs}, while the $\C$-sector is a similarly regularised version of \eqref{C-minimal}. By adjusting normalisations,  we can arrange for the seed action vertices to closely parallel our previous expressions. In fact we can get exactly \eqref{hSBD1} and \eqref{hSBD} for $B$ and $D$, while \eqref{hSAC} can be adopted by $A_1$ and $C_R$:
\be 
\label{hSA1CR}
\hS^{A_1A_1} =  2 \Delta^T_1/c_1, \quad \hS^{C_RC_R} = \Lambda^2\Delta_0/\ct_0 + 2\lambda\Lambda^4.
\ee
That leaves only the $A_2$ and $C_I$ kinetic terms and the $C_I A_2$ mixing term. These are constrained by spontaneously broken $U(1|1)$ invariance, and with our choice of normalisations can be taken to coincide with the $B${}$D$ sector:
\be 
\label{hSA2CI}
\hS^{C_IC_I} = \hS^{DD},\quad \hS^{A_2A_2} = \hS^{BB},\quad \hS^{A_2C_I}_\mu = \hS^{BD\sigma}_\mu,\quad \hS^{C_IA_2}_\mu = -\hS^{DB\sigma}_\mu.
\ee
In view of the matches \eqref{bzpt1} we already found for the kernels, we see that the flow equation for these two-point vertices coincide with \eqref{fl2} in the sense that the first equation is now for $\hS^{C_RC_R}$, the second for $\hS^{A_1A_1}$, and the last three again apply to the $B${}$D$ sector but also get copied over to the $A_2${}$C_I$ sector using the maps \eqref{bzpt1} and \eqref{hSA2CI}. The solutions for the integrated kernels are thus already given in \eqref{propCCAA} -- \eqref{propDD}, where now we should rename $\prop^{CC}$ as $\prop^{C_RC_R}$, $\prop^{AA}$ as $\prop^{A_1A_1}$ and recognise that $\prop^{A_2A_2}=\prop^{BB}$ and $\prop^{C_IC_I}=\prop^{DD}$.

\section{Manifestly gauge invariant conformal anomaly}
\label{sec:one-loop-again}

There is almost nothing left to do. Clearly equation \eqref{dS1} is replaced by
\besp
\ldl S_1 = \frac12\Tr\Big[ -\hS^{A_1A_1}\dDelta^{A_1A_1}-\hS^{A_2A_2}\dDelta^{A_2A_2} -\hS^{C_RC_R}\dDelta^{C_RC_R}-\hS^{C_IC_I}\dDelta^{C_IC_I}\\
+2\hS^{BB}\dDelta^{BB}  +2\hS^{DD}\dDelta^{DD} \Big],
\eesp
but using the  identifications of the previous section we see that the $A_2$ and $C_I$ parts just cancel half of the $B$ and $D$ parts, and thus this becomes in the old notation:
\be 
\label{dS1-Maxwell}
\ldl S_1 = \frac12\Tr\left[ -\hS^{AA}\dDelta^{AA} +\hS^{BB}\dDelta^{BB} -\hS^{CC}\dDelta^{CC} +\hS^{DD}\dDelta^{DD} \right],
\ee
\ie exactly half the result in \eqref{dS1}. Therefore we obtain half the expression in \eqref{twiceStandard}, \eqref{text-book-x2} and \eqref{beta}, \ie precisely the standard trace anomaly, however here computed by maintaining manifest gauge invariance at every stage.

\section*{Acknowledgments}
TRM thanks Roberto Percacci for helpful conversations on the Weyl anomaly, and acknowledges support from STFC through Consolidated Grant ST/L000296/1.




\bibliographystyle{hunsrt}
\bibliography{references} 

\end{document}